\documentclass{tortechrep}
\pdfoutput=1
\usepackage{url}
\usepackage{authblk}
\usepackage{hyperref}
\usepackage{comment}
\usepackage{fancyvrb}
\usepackage{fancyhdr}
\usepackage[Q=yes]{examplep}
\usepackage{marginnote}
\usepackage{graphicx}
\usepackage{caption}
\usepackage{subcaption}
\usepackage{listings}
\usepackage{xcolor}
\usepackage{fontawesome}
\usepackage{threeparttable}

\newcommand{\relaydescs}[0]{\texttt{relaydescs}}
\newcommand{\onionperf}[0]{\texttt{onionperf}}

\newcommand{\async}[0]{\texttt{async}}
\newcommand{\await}[0]{\texttt{await}}

\newcommand{\stem}[0]{\textit{stem}}
\newcommand{\asyncio}[0]{\textit{asyncio}}
\newcommand{\aiofiles}[0]{\textit{aiofiles}}
\newcommand{\aiohttp}[0]{\textit{aiohttp}}
\newcommand{\curio}[0]{\textit{curio}}
\newcommand{\asks}[0]{\textit{asks}}

\newcommand{\twisted}[0]{\textit{Twisted}}
\newcommand{\txtorcon}[0]{\textit{txtorcon}}
\newcommand{\fdesc}[0]{\textit{fdesc}}

\newcommand{\sched}[0]{\textit{sched}}
\newcommand{\schedule}[0]{\textit{schedule}}
\newcommand{\apscheduler}[0]{\textit{apscheduler}}

\newcommand{\bushel}[0]{\textit{bushel}}

\definecolor{torpurple}{HTML}{7D4698}
\definecolor{torgreen}{HTML}{68B030}
\definecolor{tordarkgreen}{HTML}{4a7f2a}
\definecolor{tordarkgrey}{HTML}{333A41}

\begin{document}
\fvset{frame=leftline,numbers=left,numbersep=2pt,gobble=4,stepnumber=1}
\VerbatimFootnotes
\graphicspath{{./images/}}


\title{Towards Modernising Data Collection and Archive for the Tor Network}
\author{Iain R. Learmonth}%
\author{Karsten Loesing}%
\affil{\vspace{-1em}\{irl,karsten\}@torproject.org \\ The Tor Project}%

\reportid{2018-12-001}
\date{December 19, 2018}
\maketitle
\tableofcontents

\begin{abstract}
  CollecTor is developed by Tor Project's Metrics Team for the purpose of
  archiving data relating to the public Tor network and applications developed
  by Tor Project.  This report distills the requirements for a prototype
  modernized replacement of the CollecTor service, and evaluates frameworks and
  libraries that are available to reduce code maintenance costs for the
  CollecTor service.
 {\renewcommand{\thefootnote}{}
  \footnotetext{This work was supported by Open Technology Fund under grant
number 1002-2017-018. Support does not imply endorsement. With thanks to Nick
Matthewson and Tim Wilson-Brown for their help in clarifying certain points in
specifications and how they are implemented in \texttt{tor}, and thanks to
Damian Johnson for his assistance in experimentation using the \stem~library.}}
\end{abstract}
\pagebreak
\section{Introduction}\label{sec:intro}%

The Tor anonymity system~\cite{tor-design} protects Internet users from
tracking, surveillance, and censorship. The Tor network is made up of thousands
of volunteer-run \emph{relays}---servers that are usually located in data
centers---distributed across the world that enable users to make private
connections to services on the Internet. Currently, the vast majority of
connections to the Tor network are made using the Tor Browser. But a growing
number of applications use the Tor network, and we expect that many more will
do so in the future.

Ongoing, robust network measurement is essential in order to respond to
censorship events, to adapt Tor Browser and other applications to respond to
changing network conditions, and to validate changes to the Tor network
software.

In the field of Internet Engineering and Privacy Enhancing Technologies it is
not common to come across large open datasets. Often this can be due
difficulties balancing utility goals with privacy risks. CAIDA, one example of
an organization that does make anonymised Internet Engineering datasets
available\footnote{An index of public datasets can be found at:
\url{https://www.caida.org/data/overview/}.}, has performed a detailed analysis
of the potential issues~\cite{caida-data-sharing}. In the field of medicine and
bio-informatics however, there has been a longer history of open data and data
re-use across studies. In one analysis, it was found that investment in the
archive and curation of open datasets had vastly greater research output
returns than solely investing in original research~\cite{Piwowar_2011}.

By collecting data about the Tor network it becomes possible to create accurate
emulations or simulations of the
network~\cite{shadow-ndss12}~\cite{ccs2013-usersrouted}~\cite{ndss13-relay-selection}.
This in turn allows for researchers to perform experiments on private testbeds
as opposed to on the public network where the experiment may harm the security
or anonymity properties of the Tor network. By collecting data over time, it is
possible to see trends in the data. For example, the blocking of Tor in China
can be identified from the data~\cite{MohajeriMoghaddam2012}. Data collection
can then also be used to validate whether or not a particular circumvention
technique is working in a particular country.

The CollecTor service fetches data from various servers in the public Tor
network and related services and makes it available to the
world\footnote{Documentation for the current implementation of the CollecTor
service can be found at: \url{https://metrics.torproject.org/collector.html}.}.
The CollecTor service provides network data collected since 2004, and has
existed in its current form as a Java application since 2010. Over time new
modules have been added to collect new data and other modules have been retired
as the services they downloaded data from no longer exist.

As the CollecTor codebase has grown, technical debt has emerged as we have added new
features without refactoring existing code. This results in it becoming increasingly
difficult to add new data sources to CollecTor as the complexity of the application
increases. Some of the requirements of CollecTor, such as concurrency or scheduling,
are common to many applications and frameworks exist implementing best
practices for these components that could be used in place of the current bespoke
implementations.

This report details the core requirements for a data collection application for
the Tor network (§\ref{sec:core-requirements}) and the specific requirements
for two modules of the application: \texttt{relaydescs} (§\ref{sec:relaydescs})
and \texttt{onionperf} (§\ref{sec:onionperf}). Library frameworks that might
be used for development of this application are then evaluated against these
requirements (§\ref{sec:frameworks}) and an initial prototype is introduced
(§\ref{sec:prototype}). Finally, next steps are identified for progressing the
development of the application (§\ref{sec:next-steps}).

\section{Core Requirements}
\label{sec:core-requirements}

\subsection{Collect}

\paragraph{Tor Relay Descriptors}

(\texttt{relaydescs})

Relays and directory authorities publish relay
descriptors, so that clients can select relays for their paths through the Tor
network. This module is discussed in more detail in §\ref{sec:relaydescs}.

\paragraph{Bridge Descriptors}

(\texttt{bridgedescs})

Bridges and the bridge authority publish bridge descriptors that are used by
censored clients to connect to the Tor network. We cannot, however, make bridge
descriptors available as we do with relay descriptors, because that would
defeat the purpose of making bridges hard to enumerate for censors. We
therefore sanitize bridge descriptors by removing all potentially identifying
information and publish sanitized versions here.

\paragraph{Bridge Pool Assignments}

(\texttt{bridgepools})

The bridge distribution service BridgeDB publishes bridge pool assignments
describing which bridges it has assigned to which distribution pool. BridgeDB
receives bridge network statuses from the bridge authority, assigns these
bridges to persistent distribution rings, and hands them out to bridge users.
BridgeDB periodically dumps the list of running bridges with information about
the rings, subrings, and file buckets to which they are assigned to a local
file. The sanitized versions of these lists containing SHA-1 hashes of bridge
fingerprints instead of the original fingerprints are available for statistical
analysis. This module has not been used since 2016, however may be reintroduced
in the future.

\paragraph{Web Server Logs}

(\texttt{webstats})

Tor's web servers, like most web servers, keep request logs for maintenance and
informational purposes. However, unlike most other web servers, Tor's web
servers use a privacy-aware log format that avoids logging too sensitive data
about their users. Also unlike most other web server logs, Tor's logs are
neither archived nor analyzed before performing a number of post-processing
steps to further reduce any privacy-sensitive parts.

\paragraph{Exit Lists}

(\texttt{exitlists})

The exit list service \textit{TorDNSEL} publishes exit lists containing the IP
addresses of relays that it found when exiting through them.

\paragraph{Torperf's and OnionPerf's Performance Data}

(\texttt{onionperf})

The performance measurement services Torperf and OnionPerf publish performance
data from making simple HTTP requests over the Tor network.  Torperf/OnionPerf
use a SOCKS client to download files of various sizes over the Tor network and
notes how long substeps take. This module is discussed in more detail in
§\ref{sec:onionperf}.

\paragraph{Future Active Measurement Modules} \faTachometer

Active measurement, from a perspective of user privacy, can be considerably
safer than passive measurement. As the Tor network continues to grow, we may
wish to expand the use of active measurement using tools such as
\textit{PATHspider}~\cite{pathspider-anrw} or
\textit{exitmap}~\cite{identifying-sybils}.

\subsection{Archive}

While it is important for clients and servers in the Tor network to have strict
validation of documents and their signatures, the CollecTor service does not
want to just drop documents that fail validation. It may be that a descriptor
is using a new format that we don't yet understand, or perhaps it is malformed
due to a bug and having the documents archived will help in debugging the
issue.

The archive should be able to verify its own integrity, ensuring that
descriptors have not been truncated or altered. It should also be possible to
determine the amount of descriptors that are missing, either through timestamps
where a descriptor/status should have been made available or by a descriptor
being referenced from another descriptor, and warn if the amount of missing
descriptors exceeds a predefined threshold.

Archiving cryptographic signatures can present challenges as the signatures
themselves use algorithms that over time will either be broken due to design
or implementation flaws, or simply due to the increase in available computing
power. A number of systems provide archive
timestamps~\cite{rfc3161}~\cite{rfc4998} where it is possible to prove that a
data object existed at a given time and so if an algorithm is considered to not
be broken at that time then the original signature can be trusted.

\subsection{Serve}

CollecTor does not only collect and archive documents, but also makes them
available to other applications. These may be other services run by Tor Metrics
such as Onionoo\footnote{This service is described at:
\url{https://metrics.torproject.org/onionoo.html}.}, or external applications
run by researchers.

For services that would like to consume all descriptors of a particular type as
they become known, CollecTor needs to make available recently obtained
descriptors. This is currently done by providing descriptors in a concatenated
form with one file per download run, however we may in the future only provide
an index to the recently downloaded descriptors to allow for applications to
fetch only the descriptors they need.

To facilitate the use of other CollecTor instances as data sources, and to
offset load generated on the network by CollecTor, a modern CollecTor may
implement parts of the Tor directory protocol version 3~\cite{dir-spec}. If
this protocol were extended to provide index functionality then the current
system of providing concatenated files for recent documents could be replaced.
This would also be of benefit for those debugging issues with the network as
individual descriptors could be easily downloaded for manual examination.

Currently the Onionoo service begins to download data from CollecTor
between :15 and :20 past the hour. If it were possible to download data sooner
than this, this would be of benefit to those monitoring the health of the Tor
network and individual relay operators as they would be able to detect problems
sooner. If CollecTor could also provide status information about the times at
which it had completed its latest download tasks, then services could consume
this in order to improve the timeliness of downloads.

For services that would like to perform historical analysis of the collected
documents, all documents must be available for download. Currently this is done
by providing monthly compressed tarballs containing the documents.

An index file that references the filenames for these concatenated files and
archives is generated to assist applications in discovering documents, but it
currently does not index the specific documents contained within the
concatenated files or tarballs.

\section{The \relaydescs~module}\label{sec:relaydescs}%

The \relaydescs~module is the primary module for data about the public Tor
network. This module collects network status votes and consensuses,
certificates, microdescriptors, and server and extra-info descriptors.
The format and purpose of each of these documents is described in version 3
of the Tor directory protocol specification~\cite{dir-spec}.

In the past, this module would also collect version 2 network statuses and
version 1 directories from the network. While we will not implement collecting
these from a live network, they should be importable via the local filesystem.

\begin{table}
\begin{center}
\begin{threeparttable}
\begin{tabular}{| l l l r r | r |}
\hline
Document & Created by\tnote{\dag} & Served by\tnote{\dag} & per hour & size ea. & size per hour \\
\hline
Detached Signature             & $A$ & $A$ & 9 & 1276B & 11.48KB \\
Status Consensus ``ns''        & $A$ & $C$ & 1 & 2.17MB & 2.17MB \\
Status Consensus ``microdesc'' & $A$ & $C$ & 1 & 2.00MB & 2.00MB \\
Status Vote                    & $A$ & $A$ & 9 & 4.34MB & 39.02MB \\
Bandwidth List\tnote{\ddag}    & $A$ & $A$ & 6 & 2.60MB & 15.60MB \\
Server Descriptor              & $R$ & $C$ & 707 & 2829B & 2.00MB \\
Extra Info Descriptor          & $R$ & $E$ & 705 & 2100B & 1.48MB \\
Microdescriptor                & $A$ & $C$ & 35 & 506B & 17.70KB \\
\hline
Total & \textemdash & \textemdash & 1473 & \textemdash & 62.30MB \\
\hline
\end{tabular}
\begin{tablenotes}
    \item[\dag] $A$ is the set of directory authorities, $E$
       is the set of extra info caches, $C$ is the set of directory caches, and
       $R$ is the set of all relays. $A \subseteq E \subseteq C \subseteq R$.
    \item[\ddag] These numbers are estimates of the numbers we will see once
       bandwidth lists are advertised from all planned bandwidth authorities.
       They were not advertised by any authority in September 2018.
  \end{tablenotes}
\end{threeparttable}
\caption{Summary of document types collected by the \relaydescs~module. Counts
per hour and average sizes are determined by the descriptors that were archived
by CollecTor for September 2018.}
\label{tbl:relaydescs-documents}
\end{center}
\end{table}

\begin{figure}
	\begin{center}
	\includegraphics[width=0.9\textwidth]{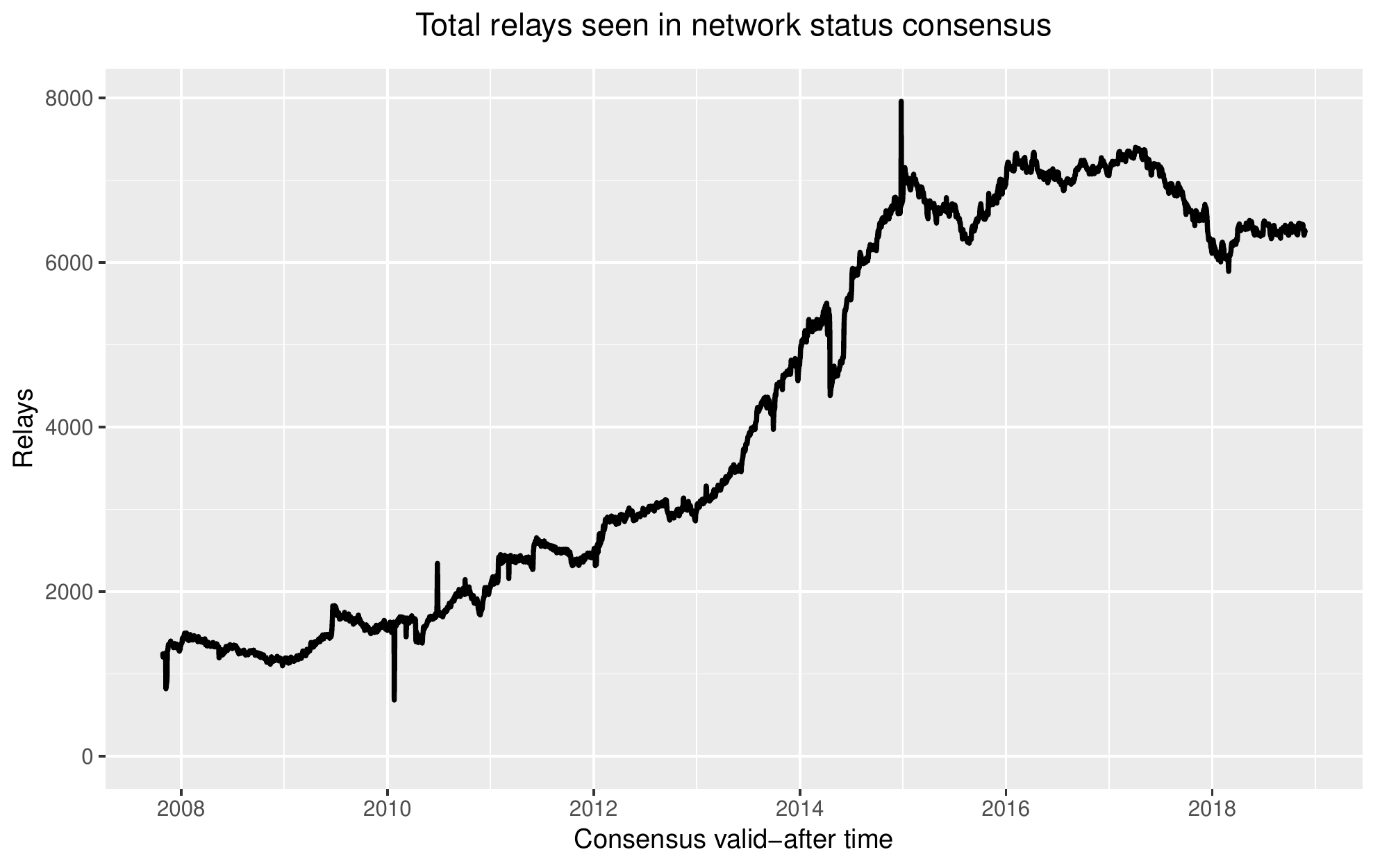}
	\end{center}
	\caption{Number of relays seen running in each consensus between
                 September 2007 and November 2018.}
	\label{fig:total-relays}
\end{figure}

\begin{figure}
	\begin{center}
		\includegraphics[width=0.9\textwidth]{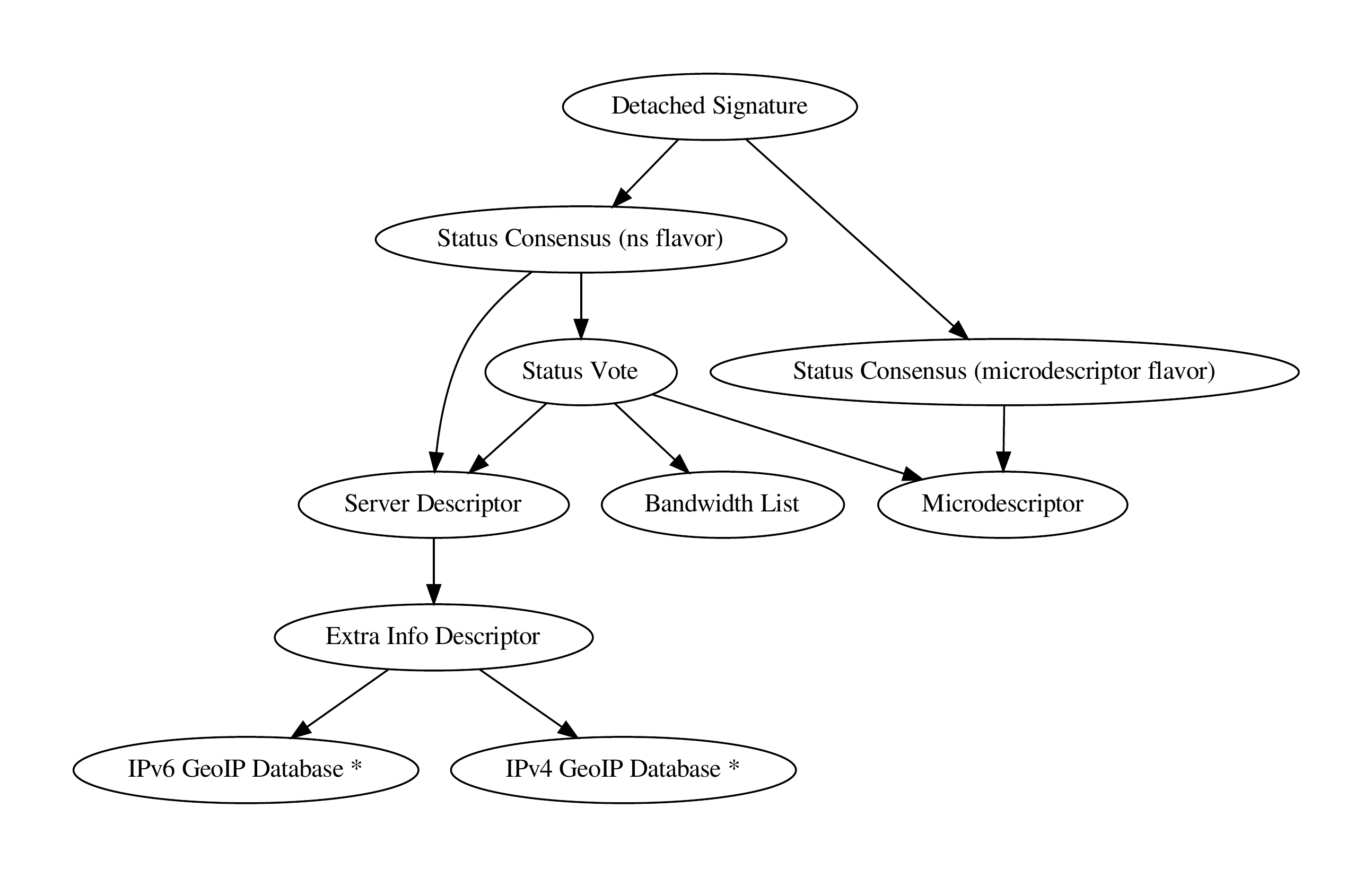} \\
        	{\footnotesize *The GeoIP databases are referenced here but not archived themselves in CollecTor.}
	\end{center}
	\caption{Document references within documents collected by the \relaydescs~module.}
	\label{fig:ref-checker}
\end{figure}

A summary of the documents collected by this module is shown in table
\ref{tbl:relaydescs-documents}. The counts and sizes of each document type are
expected to increase over time, though some more than others. The bandwidth
list document type is still under development with new features being added,
for example, while microdescriptors are intentionally minimal containing as
little information as possible for clients to still be able to function. Figure
\ref{fig:total-relays} shows the number of relays seen running in consensuses
since 2007, which directly influences the number of server, extra-info and
microdescriptors seen and also the sizes of other documents. This number has
remained relatively stable in recent years with network growth coming from
more capable relays as opposed to increased numbers of individual relays.

While most documents are served by caches, they are not instantly available
from every cache and timing must be carefully considered. References between
documents are shown in figure \ref{fig:ref-checker}. All document types can be
collected by fetching the detached signatures and recursively downloading the
referenced documents.

Unfortunately, detached signatures are only available for (typically) 5 minutes
per voting period and only from the authority that generated them\footnote{The
Tor directory protocol §3.11 does specify a URL for the detached signature that
relates to the current consensus, but this URL has not been implemented in
\texttt{tor}.}. While there are currently only two consensus flavors, there may
be more in the future and missing a detached signature means that we would not
discover it. As these documents are so tricky to get hold of, an example is
presented in appendix \ref{apx:consensus-signature}.

Without detatched signatures it is still possible to guess that a new consensus
is available when the currently known consensus is no longer ``fresh'', as
determined by the fresh-until time in the known consensus. The known consensus
flavors can then be downloaded.

\subsection{Document Sources}

This module will need to fetch data from both the network, and the local file
system. Depending on how old a descriptor is, it may be available from
different locations on the network.

Network locations include:

\begin{itemize}
\item{Directory Authorities (using version 3 of the Tor directory protocol)}
\begin{itemize}
\item{Connections might use DirPorts or tunnel over the relay's ORPort using
      the mechanism described in §2.6.1 of the Tor protocol
      specification~\cite{tor-spec}.}
\end{itemize}
\item{Directory Caches (using version 3 of the Tor directory protocol)}
\begin{itemize}
\item{As above. Additionally, directory caches that do not set
``caches-extra-info'' in their server descriptors, as described in \textsection
2.1.1 of the Tor directory protocol, may not make extra-info descriptors
available.}
\item{Future versions of CollecTor may additionally implement the Tor directory
protocol to allow for code reuse in fetching from other CollecTor instances.
This is discussed further in \textsection\ref{sec:collector-as-dirsrv}.}
\end{itemize}
\item{CollecTor instances (using CollecTor's File Structure
      Protocol~\cite{collector-protocol})}
\end{itemize}

\begin{figure}
\begin{center}
\includegraphics[width=0.9\textwidth]{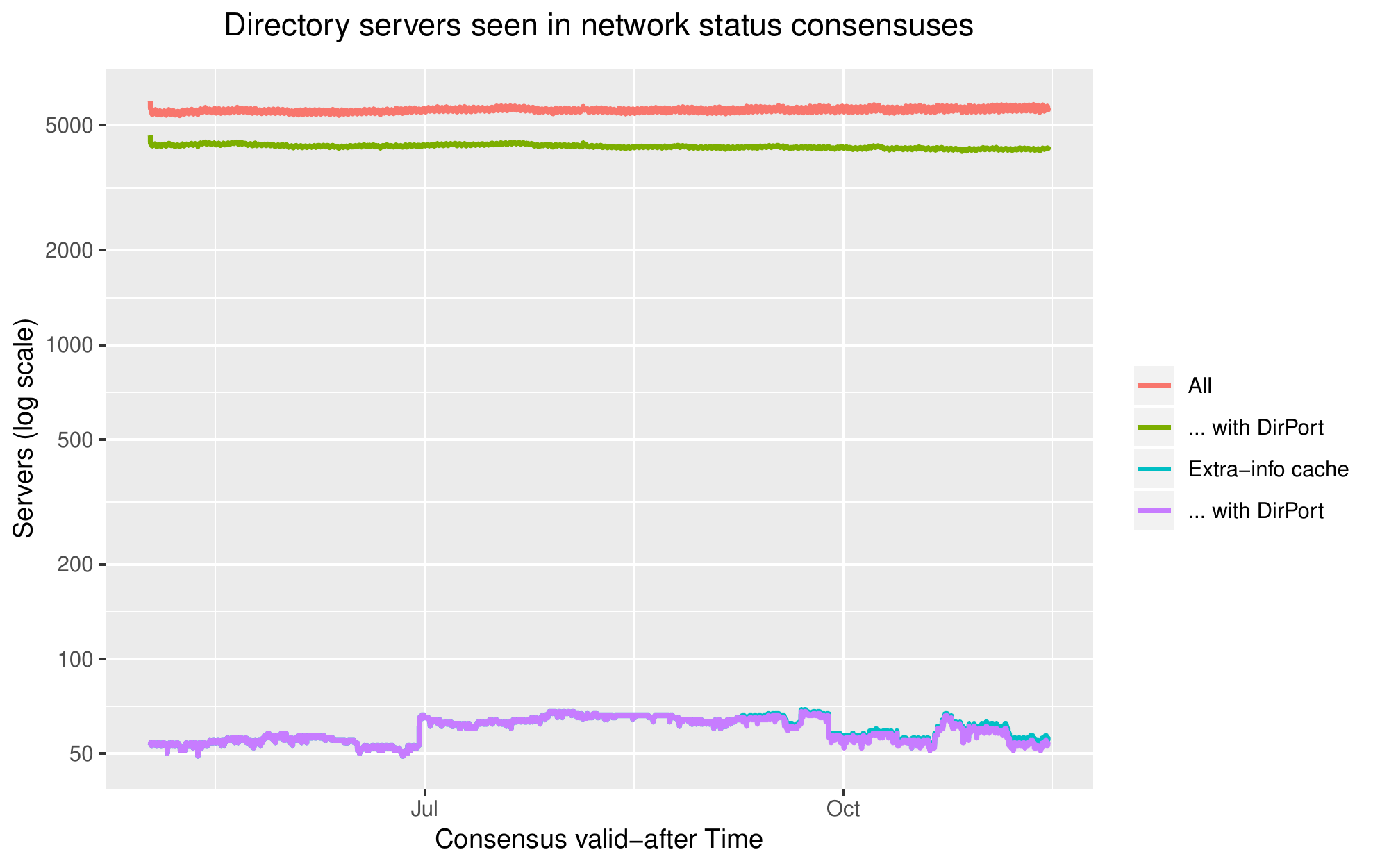}
\end{center}
\caption{Number of directory servers and extra-info caches seen running in each
consensus between May 2018 and November 2018.}
\label{fig:directory-servers}
\end{figure}

This report is written with the assumption that the Tor directory protocol
exists as-is, although conclusions from this report may influence work in
improving or extending the protocol later to improve performance, archive rate
(ratio of documents archived compared to documents missed), or to reduce
bandwidth cost.

At the time of writing there are 9 directory authorities and 2 CollecTor
instances. Figure \ref{fig:directory-servers} shows the numbers of directory
caches and extra-info caches seen in each consensus recently. In the time
period shown, there was an average of 5591 directory caches in each consensus.
There are some directory caches however that we are not currently able to use
as they do not advertise a DirPort. The \stem~library has initial support for
using ORPort tunnelling to retrieve descriptors, but it is not yet reliable.
There does not exist a Java implementation that the current implementation of
CollecTor could use to download descriptors via an ORPort. This leaves an
average of 4286 usable directory caches in each consensus.

When it comes to fetching extra-info descriptors, there are an average of 59.6
extra-info caches in each consensus. Of these, 59.2 on average advertise a
DirPort. By default extra-info descriptors are not cached by directory caches
as the descriptors are not of use to clients. If numbers are maintained at
their current levels then this should provide adequate fallback to allow
collection of descriptors if the directory authorities become unreachable.

For both directory caches and extra-info caches the trend is that the number of
caches advertising a DirPort is decreasing and so it is important to think
about how a modern CollecTor would be able to fetch via an ORPort instead.

In addition to fetching from the network, documents may also be imported from
the local file system. These formats include:

\begin{itemize}
\item{CollecTor's File Structure Protocol}
\item{Cached descriptors from a \texttt{tor} client's data directory}
\end{itemize}

\subsection{Download Scheduling}

The timing of document download tasks is determined by the valid-after
($t_{VA}$) and fresh-until ($t_{FU}$) lines found in the latest consensus.
DistSeconds ($d_{dist}$) and VoteSeconds ($d_{vote}$) are determined by the
voting-delay line in the latest consensus. $t_0$ is defined as the
time that the module is started. More information on these timings can be
found in §1.4 of the Tor directory protocol.

For all documents downloaded, the descriptors are annotated with their type and
other metadata before being saved in the archive. Each time a task is run,
the new descriptors collected should be made available either as a concatenated
file or as an index of descriptors to be downloaded by applications that would
like to consume all of a particular type of descriptor.

\paragraph{Task 0: Bootstrap}

$t = t_0$

Download the latest current consensus from a directory authority if we do not
already have one. If a download fails, try another directory authority until
all have been tried.

\paragraph{Task 1: Eager Vote Fetching}

$t = t_{FU} - d_{dist} - \frac{d_{vote}}{2}$

Download the next votes from each directory authority concurrently. During this
time the votes have not yet been computed into a consensus, but we are able to
parse the votes to get a head start on discovering new descriptors. Server
descriptors, extra-info descriptors and microdescriptors are all available to
fetch at this stage.

\paragraph{Task 2: Eager Consensus Fetching}

$t = t_{FU} - \frac{d_{dist}}{2}$

Download the detached signatures from each authority. This allows us to
discover all consensuses that have been generated.

If authorities have computed different consensuses, this is the only time at
which they can be retrieved. Archiving these alternate consensuses may prove to
be useful in debugging bugs in computing consensuses\footnote{In July 2018, a
bug occurred in the sorting of version numbers leading to 5 authorities voting
one way, and 4 voting another. Comparing the consensuses allowed the root cause
to be quickly discovered. See \url{https://bugs.torproject.org/26485} for more
information.}. A consensus requires $\frac{n}{2} + 1$ signatures, where $n$ is
the total number of known directory authorities, in order to be served via the
directory protocol as the current consensus. The voting protocol does not
preclude the existence of more than one valid consensus.

\paragraph{Task 3: Greedy Discovery} \faExclamationTriangle

While not bandwidth-friendly, directory authorities provide a method for
downloading a concatenated set of the most recent descriptors for all known
servers. This can include descriptors that have not been included in votes, but
almost certainly includes many descriptors we already know about.  The current
CollecTor implementation has support for this feature and would run this task
every 24-hours if enabled. The official Tor Metrics instances do not have this
enabled.

Download the full list of extra-info descriptors from every authority. If a
request for an authority fails, do not repeat the request. Once complete,
download the full list of server descriptors from every authority. Again, if a
request for an authority fails, do not repeat the request. The extra-info
descriptors are requested first to avoid the reference checker discovering the
extra-info descriptors from the server-descriptors and enqueueing download
tasks to retrieve them.

\begin{figure}
\begin{center}
\includegraphics[width=0.9\textwidth]{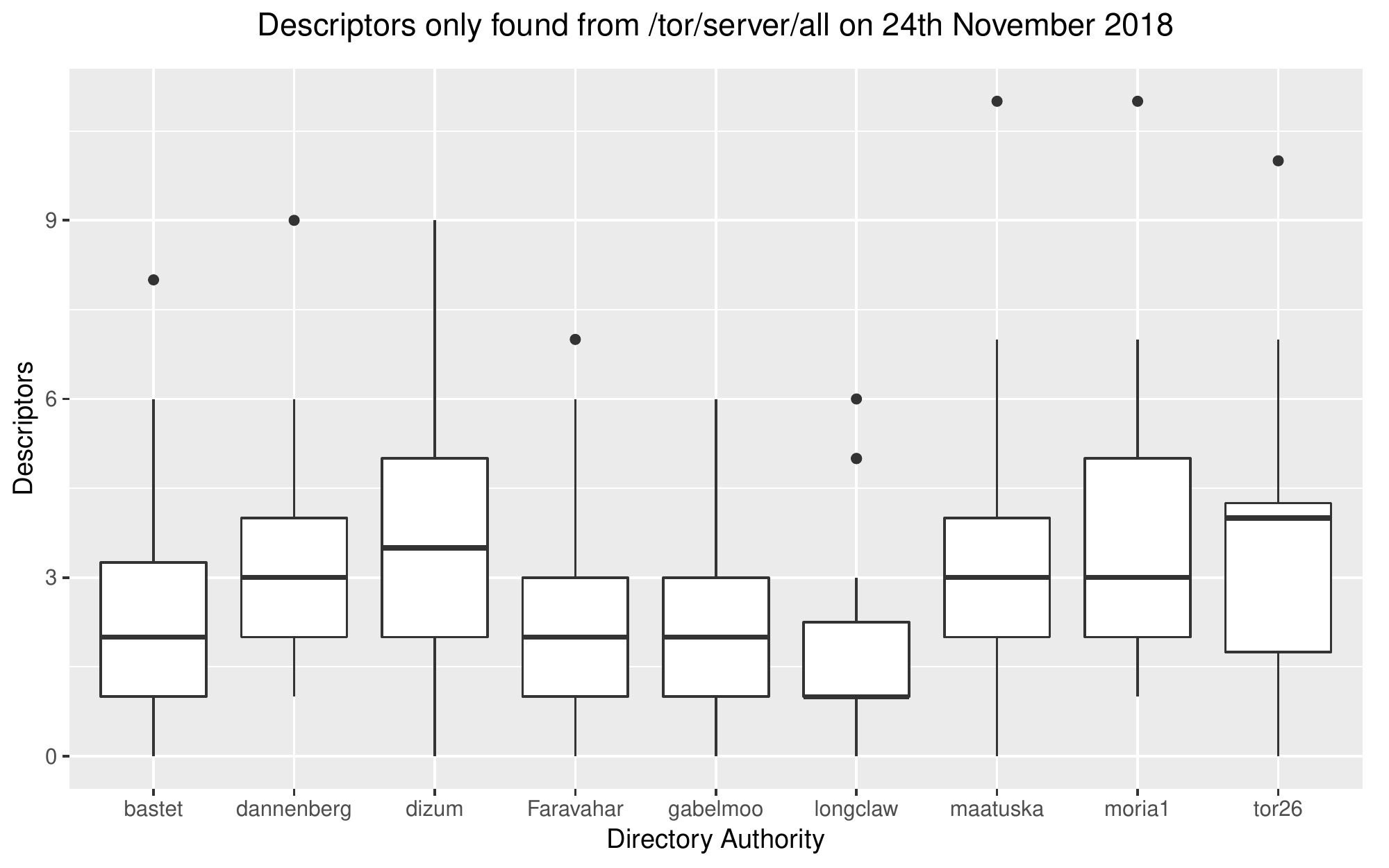}
\end{center}
\caption{Number of server descriptors that are found by requesting the URL of
all known server descriptors from directory authorities at :35 to :40 minutes
past the hour, every hour, on the 24th November 2018, that were not referenced
by the vote generated prior to, or after the download.}
\label{fig:all-descriptors-missing}
\end{figure}

An experiment performed during the preparation of this report has shown that
this is incredibly wasteful with a mean average of 2.9 descriptors discovered
by downloading all known descriptors compared to those discovered through
references in the votes prior to and after the download.  Figure
\ref{fig:all-descriptors-missing} shows the distributions across the directory
authorities. Each analysis considered only a single directory authority.

Only two instances were seen across all directory authorities in the 24 hour
period of a descriptor being available in two consecutive downloads of all
known descriptors without being referenced by the vote in between. In both
cases, the authority was ``dizum''. Upon investigation, one descriptor is for a
relay that appears to have a dynamic IP address and non-continuous uptime. The
second descriptor is for a relay that is running \texttt{tor} version 0.2.4.20,
a no longer recommended version. It is not clear why these descriptors were
retained but not used in a vote. The authority may have not found the relays to
be reachable before the vote was generated.

As each download during the experiment was approximately 17 megabytes
(uncompressed), there does not appear to be any compelling reason to enable this
feature. In order to avoid missing descriptors it would have to run every hour,
and not every 24 hours as the current implementation does.

A future extension to the Tor directory protocol may enable collecting these
descriptors by providing a URL that only serves descriptors that were not
present in the last vote. The timing to use for this request would need to be
considered unless authorities were also to make available multiple descriptors
for a single relay in this new URL instead of just the latest.

\paragraph{Task 4: Continuous Reference Checking} \faRepeat

This task runs continuously. It holds a collection of ``starting point''
documents that have been fetched by tasks 0--2. At startup, the last 3 hours of
available ``starting point'' documents will be loaded from the archive on disk
if available.

This task keeps a list of documents that have been requested since the
downloader last changed phase (described in §\ref{sec:downloader-operation}).
If a download is attempted, it won't be attempted again until the next phase.

The reference checker follows a fixed process. It first guesses at new
consensuses, consensus flavors, consensus signatures, or votes that might exist
based on the current time and fetches these, adding them to the ``starting
point'' documents. Using these:

\begin{enumerate}
\item{From each vote, bandwidth files are identified and fetched.}
\item{From each vote and consensus, server descriptors and microdescriptors are
      identified and fetched.}
\item{From each server descriptor, extra-info descriptors are identified and
      fetched.}
\end{enumerate}

When fetching server descriptors, extra-info descriptors, and microdescriptors,
these are batched to reduce the number of requests that must be made. After
each download attempt, the descriptor digests that were received are removed
from the request and it is then repeated against another server until each of
the servers available has been tried.

Following each run, starting points that are older than 3 hours are removed.

\subsection{Downloader Operation}
\label{sec:downloader-operation}

The downloader will fetch descriptors in two phases. This modifies the behavior
of the reference checker. Until a consensus is known, the downloader will
operate in phase $\alpha$. The main motivation behind the phases is to allow a
second chance for the reference checker, described in
the previous section, to locate any missing descriptors. It will also
allow for load balancing in the event that downloads are triggered for
descriptors that would now be available from directory caches.

\paragraph{Phase $\alpha$: Directory Cache Mode}

$t_{FU} - d_{dist} - \frac{d_{vote}}{2} \leqslant t < t_{VA} + \frac{t_{FU} - t_{VA}}{2}$

During this time, downloads occur in a similar manner to directory caches as
described in §4 of the Tor directory protocol. If a vote download
failed in the previous step, it must be re-attempted now. If a consensus
download failed in the previous step it must be re-attempted now. As in
phase 2, we should try to collect all available consensus signatures (or
alternate consensuses).

If a download for a particular descriptor fails, we will attempt the download
again using another authority. Within a single phase period, only one attempt
is made per authority per descriptor.

\paragraph{Phase $\beta$: Client Mode}

$t_{VA} + \frac{t_{FU} - t_{VA}}{2} \leqslant t < t_{FU} - d_{dist} - \frac{d_{vote}}{2}$

During this time, downloads occur in a similar manner to clients as described
in §5 of the Tor directory protocol. This phase gives a second
chance for descriptors that were missed earlier. This mode would also include
fetching from other known CollecTor instances via the Tor directory protocol
as discussed in the next section.

Directory caches in the network will have retrieved all the descriptors
referenced by the latest consensus by the start of this period.

\subsection{Directory Server}
\label{sec:collector-as-dirsrv}

A server, either as part of CollecTor or as a CollecTor client, implementing
the Tor directory protocol would be able to act as a directory cache. All of
the necessary documents are already available in the archive, they just need to
be returned when requested. Consensus diff\footnote{This functionality is
described in §4.5 of the Tor directory protocol.} functionality would require
some additional logic to be provided, but this functionality could also be used
to fetch consensus diffs instead of full consensuses to reduce the load on the
network created by CollecTor.

As a CollecTor instance retains descriptors for longer than the average
directory cache in the Tor network, missing descriptors could be synchronized
from other instances once they are no longer available from the caches. While
this does introduce the need to add code that serves the descriptors, it
reduces the need for alternate code to synchronize with other instances. It is
not currently possible to download individual descriptors from another
CollecTor instance.

This server would only implement a directory server and would not function as a
relay. Currently no such servers exist in the consensus but if one did, it would
be compliant with the protocol. There is a risk that such servers may provide
poor performance which would degrade client performance, and cause extra
bandwidth to be used by clients as requests may need to be retried. Directory
authorities would not perform the usual checks as there is no ORPort to use.

The Tor directory protocol previously specified a ``BadDir'' flag that could be
used to mark bad directories, indicating that clients should not attempt to use
them. This functionality was removed from \texttt{tor} in 2014\footnote{More
information about the removal of the flag can be found at:
\url{https://bugs.torproject.org/13060}.}.

\section{The \onionperf~module}
\label{sec:onionperf}

In comparison to the \texttt{relaydescs} module, this module is a lot simpler.
OnionPerf\footnote{The source code and documentation for OnionPerf can be found
at: \url{https://github.com/robgjansen/onionperf}.}~uses multiple processes and
threads to download random data through Tor while tracking the performance of
those downloads. The data is served and fetched on localhost using two TGen
(traffic generator) processes, and is transferred through Tor using Tor client
processes and an ephemeral Tor Onion Service. Tor control information and TGen
performance statistics are logged to disk, analyzed once per day to produce a
json stats database and files that use the Torperf results format, and can
later be used to visualize changes in Tor client performance over time.

\subsection{Document Sources}

This module collects the Torperf formatted results files from OnionPerf
instances, of which Tor Metrics currently has 3. One result file is produced at
midnight each day for each of the file sizes configured to test with. There are
three file sizes used for measurements: 50 KiB, 1MiB and 5MiB. This means that
we collect $3 \times 3 = 9$ results files each day. The size of the downloads
are chosen probabilistically and so it is not easy to predict the sizes of
each file. In September 2018, a total of 24 MB of results were collected.

\subsection{Download Scheduling}

Each day the scheduler should start downloads of the results from the previous
day. There are no other sources available, except perhaps other CollecTor
instances, for the files and so if a file is unable to be retrieved due to a
permanent error it should not be reattempted.

\section{Frameworks Evaluated}\label{sec:frameworks}%

All of the evaluated frameworks use the Python language, initially targeting
version 3.7\footnote{Should it be necessary to deploy any replacement service
before the next Debian release, it would be possible to use lower-level
mechanisms to recreate the Python 3.7 language features we use, but this would
mean additional code complexity and maintenance costs, which we are trying to
reduce.}.  Tor Metrics runs its services on Debian stable systems. The next
Debian release, Debian 10 ``buster'', is expected mid-2019 and will include
Python 3.7 or later.

The CollecTor service is still well within the limits for operation on a single
machine and so while distributed frameworks such as Apache Beam do offer
scalability, in this case it is unnecessary and would lead to additional
complexity in the codebase. There are four main areas in which we would like to
re-use an existing framework: descriptor parsing, concurrency, scheduling and
plugin architecture.  Each of the frameworks is evaluated for its applicability
to the application and its ability to reduce software development and
maintenance costs for the CollecTor application.

\subsection{Document Parsing}

CollecTor needs to work with many document formats that are specific to the Tor
ecosystem. The current Java implementation of CollecTor uses
metrics-lib\footnote{The documentation can be found at:
\url{https://metrics.torproject.org/metrics-lib/}.} which is primarily
maintained by Tor Metrics for the use of applications developed by Tor Metrics.

\paragraph{stem}~

\stem\footnote{The documentation can be found at:
\url{https://stem.torproject.org/}.}~is a Python library for parsing
Tor-specific data formats, and for interacting with remote Tor servers (i.e.
directory servers). It does not support all the current formats supported by
metrics-lib although this support can be added. The library is also used as
part of Tor Project's \texttt{nyx} application and as part of the test-suite
for \texttt{tor} which means that it is being exercised by more developers than
just the Tor Metrics team and hopefully allows for issues to be quickly
discovered and fixed.

\subsection{Concurrency}

The vast majority of the work performed by CollecTor is I/O bound. That is to
say that the time it takes to complete a task is determined principally by the
period spent waiting for I/O (network or disk) operations, to be
completed. When fetching server descriptors, extra-info descriptors, or
microdescriptors, there will typically be thousands of descriptors to fetch
before moving on to the next stage. Downloads of descriptors of the same type
do not depend on each other and so are candidates for concurrent execution.

The current Java implementation of CollecTor uses
\texttt{java.util.concurrent}\footnote{The documentation can be found at:
\url{https://docs.oracle.com/javase/8/docs/api/java/util/concurrent/package-summary.html}.}~
to provide concurrency, with the tasks running in threads. All synchronisation
between tasks must be performed manually.

\paragraph{asyncio}~

\asyncio\footnote{The documentation can be found at:
\url{https://docs.python.org/3.7/library/asyncio.html}.}~\cite{PEP3156} is a
framework for asynchronous programming in Python. Coroutines declared with
\async/\await~syntax~\cite{PEP492} is the preferred way of
writing \asyncio~applications. While callbacks are possible, they are not used
explicitly in practice. Future objects, which represent an eventual result of
an asynchronous operation, are used to bridge low-level callback-based code
with high-level \async/\await~code.

Other language features, for example the ability to delegate to a
subgenerator~\cite{PEP380}, allow for concurrent programming while writing in a
sequential fashion. Parallel computing using threads is hard because of race
conditions. \asyncio~ is explicit about where the event loop may take control
of the program. This reduces mental load for developers as resulting programs
are easier to follow, which should help to reduce development and maintenance
costs directly.

While the \stem~library does not have native support for \asyncio~it does have
support for asynchronous requests and a simple wrapper can be written to allow
integration. An example is shown in listing \ref{lst:stem-asyncio-wrapper}.

\asyncio~is part of the Python standard library. It may still be quite new but it has momentum.

\begin{figure}
\lstset{caption={Python \asyncio~wrapper for \stem~to download the latest consensus},label=lst:stem-asyncio-wrapper}
\begin{lstlisting}
async def fetch_consensus():
    """
    Returns the latest consensus.
    """
    loop = asyncio.get_running_loop()
    query = stem.descriptor.remote.Query("/tor/status-vote/current/consensus",
        document_handler=stem.descriptor.DocumentHandler.DOCUMENT)
    result = await loop.run_in_executor(None, query.run)
    for consensus in result:
        return consensus
\end{lstlisting}
\end{figure}

For modules like the \onionperf~module, that require only to fetch data from
a remote HTTP server, the \aiohttp\footnote{The documentation can be found at:
\url{https://docs.aiohttp.org/}.}~library provides an \asyncio-compatible
asynchronous HTTP client. This library also includes web server functionality
that could be used to serve archived documents.

Local file I/O is blocking, and cannot easily and portably made asynchronous.
While there has been efforts to bring asynchronous file I/O to POSIX and Linux
it does not seem to have been adopted by developers. To avoid file I/O blocking
execution, we can make use of the \aiofiles\footnote{The source code and
documentation can be found at:
\url{https://github.com/Tinche/aiofiles}.}~library which provides an object
with an API identical to an ordinary file. The asynchronous I/O is provided by
delegating I/O operations to a thread pool.

\paragraph{curio}~

\begin{figure}
\lstset{caption={Python \curio~wrapper for \stem~to download the latest consensus},label=lst:stem-curio-wrapper}
\begin{lstlisting}
async def fetch_consensus():
    """
    Returns the latest consensus.
    """
    query = stem.descriptor.remote.Query("/tor/status-vote/current/consensus",
        document_handler=stem.descriptor.DocumentHandler.DOCUMENT)
    result = await curio.run_in_thread(query.run)
    for consensus in result:
        return consensus
\end{lstlisting}
\end{figure}

\curio\footnote{The documentation can be found at:
\url{https://curio.readthedocs.io/}.}~is a library of building blocks for
performing concurrent I/O and common system programming tasks such as launching
subprocesses, working with files, and farming work out to thread and process
pools. It uses Python coroutines and the explicit \async/\await~syntax but does
not use \asyncio.

While \curio~is not part of the Python standard library it does not have any
third-party dependencies. It is quite low-level however and so there would
likely be work in building enough infrastructure on top of it to handle the
tasks we would like to perform.

While the \stem~library does not have native support for \curio, it is again
simple to create a wrapper for integration.  An example is shown in listing
\ref{lst:stem-curio-wrapper}. This is very similar to the \asyncio~wrapper as
it is using the same concepts.

There is no HTTP support available from \curio, nor a recommended HTTP library
to use. The \asks\footnote{The documentation can be found at:
\url{https://asks.readthedocs.io/}.}~library provides a \curio-compatible HTTP
client but does not implement a server.

\curio~does provide support for asynchronous file operations. Like
\aiofiles~it uses threads, however this may change in the future. The use of
threads is noted as an implementation detail in the documentation which may
indicate that this would change in the future to be the most optimised
mechanism for the platform that is in use.

One strong feature of \curio~is that it recognises that asynchronous
programming is still new to Python and provides primitives, such as
\textit{curio.UniversalQueue}, that allow for communication between \async~tasks
and threads. The \asyncio~counterpart, \textit{asyncio.Queue}, will only permit
communication between \async~tasks. This allows for transition between legacy
libraries and those that support the new language features, however this is a
fresh implementation of the application using Python so we do not have problems
with legacy dependencies.

\paragraph{Twisted}~

The \twisted~framework for Python is very mature event-driven framework and has
support for a large number of network protocols. It does not have support for
Tor's directory protocol although this could be built on top of a \twisted~HTTP
client for DirPort. For ORPort usage a minimal implementation of the Tor
protocol would be required using \twisted~for the directory protocol client to
use, which is a non-trivial piece of work.

There exists a library for Tor's control protocol~\cite{control-spec},
\txtorcon\footnote{The documentation can be found at:
\url{https://txtorcon.readthedocs.io/en/latest/}.}, however this application
is primarily concerned with the directory protocol and the documentation for
\txtorcon~points users towards \stem~for this.

\begin{figure}
\lstset{caption={Python \twisted~wrapper for \stem~to download the latest consensus},label=lst:stem-twisted-wrapper}
\begin{lstlisting}
def fetch_consensus():
    """
    Returns the latest consensus.
    """
    query = stem.descriptor.remote.Query("/tor/status-vote/current/consensus",
        document_handler=stem.descriptor.DocumentHandler.DOCUMENT)
    query.run()
    for consensus in result:
        return consensus

def handle_consensus(consensus):
    """
    Do something with the consensus.
    """
    ...

t = reactor.deferToThread(fetch_consensus)
t.addCallback(handle_consensus)
reactor.run()
\end{lstlisting}
\end{figure}

Wrapping \stem~is again possible as shown in listing
\ref{lst:stem-twisted-wrapper}, but the reactor pattern used by \twisted~makes
using this more complicated. It is not possible to pause the execution of the
calling function as with the \async/\await~syntax and so a callback must be
used. This pattern inverts the flow of control and so makes the code more
difficult to debug than the pseudo-procedural pattern made possible by
\async/\await~and other related language features.

Non-blocking file I/O is provided by the \fdesc\footnote{The documentation can
be found at:
\url{https://twistedmatrix.com/documents/current/api/twisted.internet.fdesc.html}.}~module
but the API for this is very limited. It also operates directly on file
descriptors and does not provide a complete abstraction for files.

\subsection{Scheduling}

Each module needs to download documents on a schedule. Timing can be very
important as there may only be a small window in which documents are available
for download before they are discarded. The current Java implementation uses
\texttt{java.util.concurrent} which provides basic scheduling functionality.

The article~\cite{rethinking-cron} that inspired \schedule, evaluated below,
describes a wishlist for a scheduling solution. First, it must have a powerful
and human-friendly syntax. This is particularly important for CollecTor as
there will be a number of scheduled tasks to perform per module and it is
important that mistakes are not made. To correctly implement the Tor directory
protocol specification, some times must be calculated based on values found in
the latest consensus and cannot simply be declared with a crontab-like syntax.

Testing is also important as a means to reduce development costs. By being able
to easily validate the scheduling of tasks, and also test tasks themselves in
an environment that does not differ from the environment used for scheduled
execution, it is possible to catch bugs before software changes are deployed.

The operation of the scheduler must be clear to ensure that tasks are running
correctly and to assist in any debugging. This can be achieved by having good
visibility into the scheduler through logging and performing as little work in
the scheduler as possible with all heavy lifting being performed by individual
tasks.

\paragraph{sched}~

The \sched\footnote{The documentation can be found at:
\url{https://docs.python.org/3.7/library/sched.html}.}~module, part of the
standard library for Python 3.7, provides a general purpose scheduler. While it
is nice to not have external dependencies, it operates on a monotonic clock
and does not understand UTC time on which the directory authorities, and other
services that CollecTor must interact with, operate. It also provides no
facility for recurring tasks or for scheduling tasks to run at a specific time,
only to run tasks once after a delay.

\paragraph{schedule}~

\schedule\footnote{The documentation and source code can be found at:
\url{https://github.com/dbader/schedule}.}~is an in-process scheduler for
periodic jobs that uses the builder pattern for configuration. The syntax is
easy to understand and so should reduce mistakes. It supports scheduling tasks
to run at periodic intervals, or at fixed times. It does not support scheduling
a task to run only once without modifying the task to cancel its schedule after
its execution.

It expects that programs will either have thread-safe tasks or that the
developer will take care of ensuring safe execution of the tasks. Listing
\ref{lst:schedule-asyncio-wrapper} shows how a wrapper might be used to run
\asyncio~tasks using \schedule.

The current maintainer has indicated the he does not have the time to properly
maintain this package and is seeking to bring on a co-maintainer\footnote{More
discussion may be found at the GitHub issue:
\url{https://github.com/dbader/schedule/issues/219}.}~which indicates a risk
that if this library is used, Tor Metrics may become the de-facto maintainers
of it.

\begin{figure}
\lstset{caption={\schedule~wrapper for \asyncio~tasks},label=lst:schedule-asyncio-wrapper}
\begin{lstlisting}
import asyncio
import time
import schedule
from threading import Thread

loop = asyncio.new_event_loop()

def f(loop):
    asyncio.set_event_loop(loop)
    loop.run_forever()

t = Thread(target=f, args=(loop,))
t.start()

def run_async(job_coro):
    job_task = job_coro()
    loop.call_soon_threadsafe(asyncio.async, job_task)

async def job():
    await asyncio.sleep(1)
    print('Hello, world!')

schedule.every(10).seconds.do(run_async, job)

while 1:
    schedule.run_pending()
    time.sleep(1)
\end{lstlisting}
\end{figure}

\paragraph{Advanced Python Scheduler}~

Advanced Python Scheduler\footnote{The documentation can be found at:
\url{https://apscheduler.readthedocs.io/}.}, also known as \apscheduler, is an
in-process scheduler for periodic jobs that provides an object to add jobs to
at runtime, or permits for scheduled tasks to be added by using a decorator.

It supports scheduling tasks to run at periodic intervals, at fixed times, and
also for a single execution at a fixed time or interval. Jobs can be stored
persistently on disk, and \apscheduler~will check for misfired jobs (where the
job was unable to be executed at the desired time) and run the job immediately
if it is configured to do so.

By default, only one instance of each job is allowed to be run at the same
time. This means that if the job is about to be run but the previous run hasn't
finished yet, then the latest run is considered a misfire. It is possible to
set the maximum number of instances for a particular job that the scheduler
will let run concurrently.

\apscheduler~provides a scheduler that runs on an \asyncio~event loop that can
run jobs based on native coroutines using the \async/\await~syntax. It also
provides a scheduler that runs on a \twisted~reactor that uses the reactor's
thread pool to execute jobs.

\subsection{Plugin Architecture}

By building CollecTor as an extensible application, it allows easy addition of
new data sources in the future. It allows for both Tor Metrics and third-party
developers to easily enhance your software in a way that is loosely coupled:
only the plugin API is required to remain stable. This extensibility is
achieved through the definition of one or more APIs and a mechanism for
collecting code plugins which implement this API to provide some additional
functionality.

\paragraph{twisted.plugin}~

This is a component of Twisted, which was previously evaluated for its
concurrency features, but can also be used as a standalone module. It has a
dependency on \textit{zope.interface} which is used to define interfaces for
plugins.

It allows new plugins to be discovered flexibly. For example, plugins can be
loaded and saved when a program is first run, or re-discovered each time the
program starts up, or they can be polled for repeatedly at runtime (allowing
the discovery of new plugins installed after the program has started).

Overall this is quite a heavy module and the complexity in its dependencies may
cause more trouble than the benefits it brings are worth.

\paragraph{straight.plugin}~

This module is quite light but also does not have any interface mechanism.
Instead, plugins are found from a namespace and can be identified by a parent
class. Through namespace packages, plugins can be split up into separate
codebases, even managed by different teams, as long as they all implement the
same base API.

If interfaces are required the mechanisms in \texttt{collections.abc}, part of
the standard library, may provide a suitable implementation.

\subsection{Discussion}

Whatever other frameworks are used, \stem~is the only viable choice for
descriptor parsing if targeting Python 3.7. Fortunately it is well maintained
and is a mature stable library. During the course of preparing this report, a
number of features were included in \stem~to assist in experimentation
including:

\begin{itemize}
	\item{Parsing descriptors from a byte-array (\href{https://bugs.torproject.org/28450}{\#28450})}
	\item{Parsing of detached signatures (\href{https://bugs.torproject.org/28495}{\#28495})}
	\item{Generating digests for extra-info descriptors (\href{https://bugs.torproject.org/28398}{\#28398})}
	\item{Generating digests for votes and consensuses (\href{https://bugs.torproject.org/28398}{\#28398})}
	\item{Generating digests for microdescriptors (\href{https://bugs.torproject.org/28398}{\#28398})}
\end{itemize}

While potential authors of libraries that would compete with \stem~should not
be discouraged from implementing alternatives, \stem~does fill all of the
requirements of the CollecTor application for the parsing of descriptors.

For concurrency, the \asyncio~framework appears to be the best choice. Moving
away from a threading model to an asynchronous model it provides all the
functionality required for the CollecTor service requirements. \curio~would
also have been a viable option however it has a smaller community than
\asyncio~and so less library code is readily available for reuse. There does
not appear to be a compelling advantage to using \twisted~over the more modern
frameworks that make use of new language features such as the
\async/\await~syntax despite its maturity.

In the evaluation of these frameworks it became clear that performing file I/O
operations in an asynchronous way is not simple. The \asyncio~framework
abstracts the complexity by delegating the blocking operations to a thread pool
however in the longer term we may wish to explore other storage options.

For scheduling, Advanced Python Scheduler is the only library evaluated that
fits the requirements for the CollecTor service. The native support for the
\asyncio~event loop means that no custom wrappers will be required. Both
\sched~and \schedule~would be useful for other tasks, but for CollecTor are too
minimal.

For the plugin architecture, \textit{straight.plugin} is the clear choice as
the Twisted module is very heavy in comparison without providing any clear
advantages.

\section{Prototype Implementation}\label{sec:prototype}%

\begin{figure}
	\begin{center}
		\includegraphics[width=0.4\textwidth]{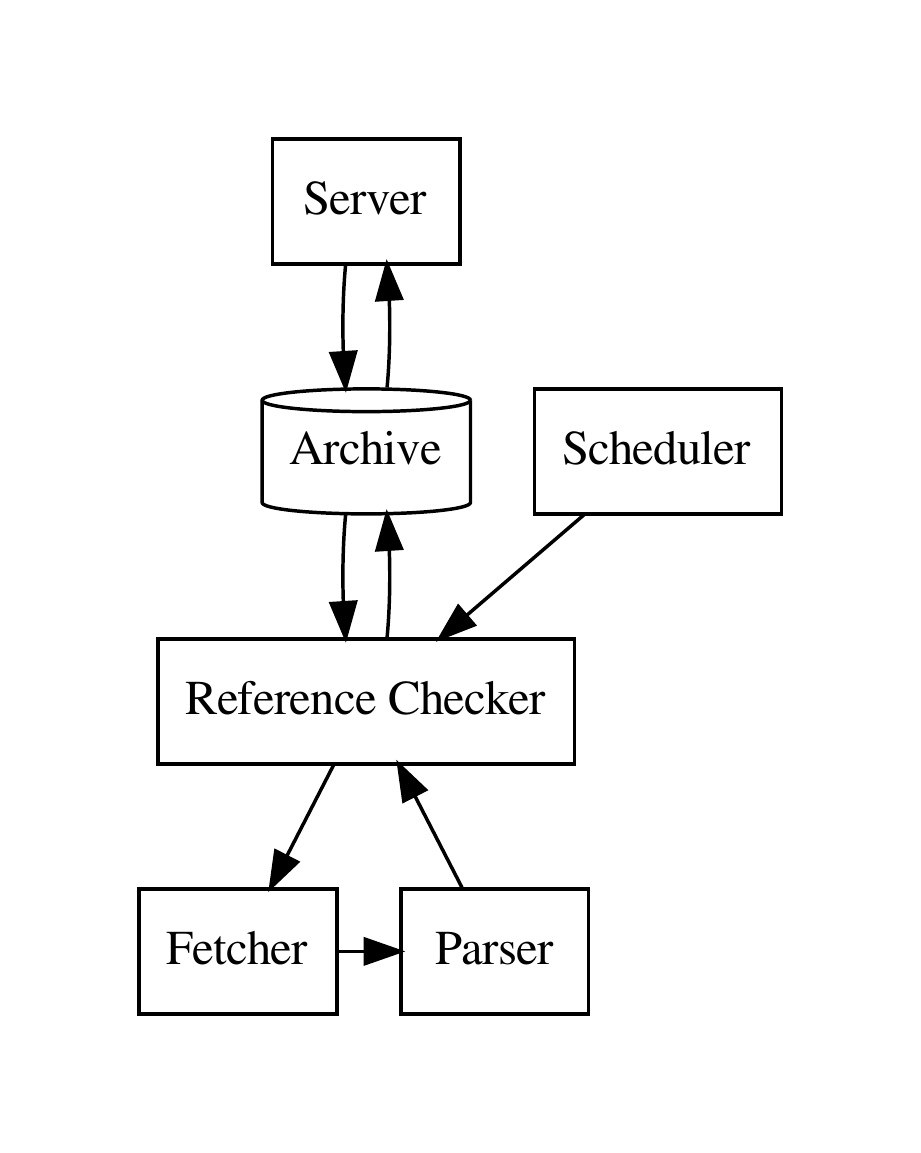}
	\end{center}
	\caption{Overview of the architecture for the next-generation CollecTor.}
	\label{fig:architecture}
\end{figure}

A prototype of an application implementing the requirements described in
§\ref{sec:relaydescs} has been implemented. This prototype is known as
\bushel~and the source code and documentation can be found online\footnote{The
source code can be found at \url{https://github.com/irl/bushel} and the
documentation at \url{https://irl.github.io/bushel}.}.

The prototype makes use of \asyncio~for asynchronous I/O. Where using the
\stem~library, any calls that would have blocked are delegated to an executor,
currently a \texttt{concurrent.futures.ThreadPoolExecutor}.

The primary functionality of the \relaydescs~module is implemented in the
\texttt{DirectoryScraper}. This has functionality for recursively discovering
documents that should be archived.

The \texttt{DirectoryCache} provides an abstraction layer that forwards
requests to a \texttt{DirectoryArchive} or a \texttt{DirectoryDownloader}
instance. When scraping the directory documents are requested from the
\texttt{DirectoryCache}. If they are not found, and a download is successful,
they are stored in the archive as a side-effect.

The \texttt{DirectoryArchive} provides methods to retrieve descriptors that
have been archived in the local file system. When parsing a consensus there are
roughly 6500 server descriptors referenced, and even more for a vote, which is
well above the default number of maximum file descriptors for a
process\footnote{The default maximum file descriptors per process is 1024 on
Debian 9 systems, and remains unchanged in Debian 10 at time of writing.}. To
prevent unbounded use of file descriptors, an \texttt{asyncio.BoundedSemaphore}
is used to limit concurrency.

\section{Next Steps}\label{sec:next-steps}%

\begin{itemize}

\item{Based on the experience of implementing this prototype, a draft plugin
API has been specified in appendix \ref{apx:plugin-api}. The prototype will
require some refactoring to fit this API and then enable the implementation of
the requirements set out in §\ref{sec:onionperf}.}
\item{Currently the prototype runs only as a command-line tool and not as a
service with an in-process scheduler. The scheduler would need to be integrated
to the prototype before it could be deployed.}
\item{The API may still require new functions or tweaks to existing functions
and would need to be formalised in the documentation. Before it can be
considered complete an assessment of suitability for each of the current
CollecTor modules would need to be performed.}
\item{In order to improve the archive rate for detached signatures, which
currently must be collected during a strict 5 minute interval, it would be
useful to have the missing URL that publishes the detached signature for the
current consensus implemented in \texttt{tor}. The Tor directory protocol could
further be extended to support retrieval of recent consensuses, votes and
detached signatures and not just those for the current and next periods.}
\item{For server descriptors our archive rate will not be 100\% due to relays
uploading new descriptors twice between CollecTor polling the directory
authorities. One possible solution to this would be to provide a URL to
retrieve all known descriptors, not just the most recent. URLs could be
provided to limit the descriptors to only those learned within a given time
period to help reduce duplicated downloads while maintaining a high archive
rate.}
\item{The archive rate will need to be monitored, and to define thresholds for
warning the service operators the current CollecTor archives should be analyzed
to find a baseline. Any replacement needs to at least maintain this baseline,
if not improve on it.}
\item{Synchronization between CollecTor instances has not yet been considered.
While the current CollecTor implementation supports this through the CollecTor
client interface, it is suboptimal in terms of bandwidth usage and an improved
design may help both synchronization and for general client usage.}
\item{Currently there are no efforts to provide trusted timestamps for
documents containing signatures that are archived by CollecTor, but in the
future we could look into providing this service.}
\item{Finally, alternatives for document storage may be considered. This report
assumed that a new implementation would continue to implement the CollecTor
File Structure Protocol however this is not a strict requirement for the
internal storage. Using the same structure on top of ZFS, using a relational
database, or using an object store could provide better performance and reduce
application complexity with some tasks delegated to the storage provider.}
\end{itemize}

\pagebreak
\appendix
\addtocontents{toc}{\protect\setcounter{tocdepth}{0}}

\section{Sample Detached Consensus Signature}\label{apx:consensus-signature}%

\lstset{caption={Sample detached consensus signature},label=lst:consensus-signature}
\begin{lstlisting}
consensus-digest 1CBD322788FFC841B0DB701C2942EE5750617CFF
valid-after 2018-11-15 19:00:00
fresh-until 2018-11-15 20:00:00
valid-until 2018-11-15 22:00:00
additional-digest microdesc sha256 476993E797C51682E95ACEED12B2DD21588847E8E2FF7C49291E64207D8FED53
additional-signature microdesc sha256 D586D18309DED4CD6D57C18FDB97EFA96D330566 8A45BACC94A6023A90C24FBCD10520C1741828F7
-----BEGIN SIGNATURE-----
1c/vHIqlqdhS8HR+Lps3Tk+VHeJaQ5lL/NxIkARDpVMLhv6fHxCNGlXrKvd9S5KR
MvOzblmrVt3TV/iJTvOmMwHuziRjzrZeHpeeK81zQ/z6QGvheooaxa8jsYuANgA0
GK4agnsCI4JTKz/47SGpIDjY3VtXbns58TUPYHHUQY82khLqWvj1nL5djWdnnm9l
yyU4od4mv6JJz9XdCNN+qDTzEA0QE10Y0lUV+K2Ipqplrb/zd9pzJS9GUf82cNOj
GYLvBMzuSr/aL0UIeQgiI0BRDw2MPqXd/KA04dOFCiqnDhKqh0PR6SMD3ulgxxhs
R0du41KYQC/eDqeRhxZF4g==
-----END SIGNATURE-----
directory-signature D586D18309DED4CD6D57C18FDB97EFA96D330566 8A45BACC94A6023A90C24FBCD10520C1741828F7
-----BEGIN SIGNATURE-----
ITaD0D5CmuobYi3G5LbuWmbIe5Vpt3o+5d1XOtKaBhRxmC10c9WWMXCVJ7K6Ezb5
dzX6CsEKpop1+V8eqPRTyAZ7H4VvxNS5j6yPsgrMlahgQjcaOpxZY8p+dmzEluPe
E45/+qlXoNfxwF4jv1t1+NLM0jIJRwHErNgJXzFRZ/q/MUZxn/LuN68mcBqzdLD4
L/D9bKNmvIAkcfTedk0x/zmwaXNMV6N9kN3kmUqeAvFLNOM/oP46ktj+B5Ch/2et
lFy4MEf1iHXKiLzq2uuCkMN2pfVtmga8j/BHE47ne5paMHnDwaTrEmBM2ws8n4mK
E/RAIUlD8COyEUImjcns6w==
-----END SIGNATURE-----
\end{lstlisting}

\pagebreak
\section{Initial Plugin API}\label{apx:plugin-api}%

The following documents a draft API to be implemented by plugins.
These functions will be called by the reference checker. While plugins
may keep state internally, it is expected that any state they do keep
is not required to be persistent.

The latest version of this API will be found at
\url{https://irl.github.io/bushel/plugins.html}.

\begin{verbatim}
class DocumentIdentifier(doctype, subject, datetime, digests):

   Represents a document that is expected to exist.

   **Attributes:**

   doctype

      The "type" of the document.

   subject

      The subject of the document. This is usually a string containing
      an opaque identifier. Examples include the fingerprint of a
      relay for a server descriptor, or the hostname of an OnionPerf
      vantage point.

   datetime

      A "datetime" related to the document. The exact meaning of this
      will be document dependent. Example include the published time
      for a server descriptor, or the valid-after time for a network
      status consensus.

   digests

      A "dict" containing mappings of "DigestHash" to "tuple"s. Each
      tuple contains a "str" representation of the digest
      and a "stem.descriptor.DigestEncoding".
\end{verbatim}
\pagebreak
\begin{verbatim}
class ExamplePlugin:

   An example plugin for bushel.

   expectations()

      Returns:
         A "list" of "DocumentIdentifier" for documents that are
         expected to be available for fetching.

   fetch(docid)

      Fetches a document from a remote location.

      Parameters:
         **docid** (*DocumentIdentifier*) – Identifier for the
         document to be fetched.

   parse(document)

      Parses a retrieved document for any documents that are
      referenced and should be fetched.

      Parameters:
         **document** (*Document*) – A retrieved document.

      Returns:
         A "list" of "DocumentIdentifier" for documents that are
         expected to be available for fetching.
\end{verbatim}

\end{document}